# Acceptance-Test-Driven Evaluation Protocols for Business-Centric LLM Systems

Eric Liang

Oracle

zixuan.liang@oracle.com

**Abstract.** Large language model (LLM) applications are increasingly expected to satisfy deterministic institutional requirements while relying on probabilistic generative components. This mismatch makes ordinary post-hoc benchmarking insufficient for systems that must be safe, reliable, auditable, and economically useful. This paper contributes an evaluation-protocol extension for operational LLM systems grounded in acceptance-test-driven development, safety engineering, and business-centric validation. The extension translates stakeholder goals into executable behavioral contracts, release gates, monitoring signals, and evidence artifacts before prompt, model, retrieval, or agent changes are accepted. It adapts the red-green-refactor discipline of test-driven development to a red-train-green lifecycle: first define failing acceptance tests for desired behavior, then improve the LLM system through prompt changes, retrieval design, fine-tuning, guardrails, or data augmentation, and finally release only when multidimensional gates are satisfied. The contribution is a governance-oriented metric stack, reference architecture, and empirical protocol for comparing acceptance-test-driven LLM development against prompt-first and benchmark-after workflows.

## 1. Introduction

LLM systems have moved from demonstrations into operational settings such as customer service, clinical intake, enterprise search, legal drafting, education, software engineering, and decision support. These systems are expected to be useful under diverse language, intent, and context conditions, yet they can hallucinate, over-refuse, leak sensitive information, follow malicious instructions, or silently regress when the model, prompt, retrieval corpus, or tool environment changes. The central engineering problem is therefore not merely whether an LLM can produce a plausible answer once, but whether the surrounding system can repeatedly satisfy explicit requirements under uncertainty.

Classical test-driven development (TDD) addresses a related discipline problem by writing tests before implementation and using the red-green-refactor cycle as a requirements-to-code feedback loop [1]. Acceptance test-driven development extends this idea toward stakeholder-visible behavior. Farago described acceptance-test-driven LLM development, and Parupally subsequently introduced ATDLLMD as a framework that combines agile practices, data-centric methods, safety engineering, CPMAI-style process structure, and LM-Eval tooling for business-centric LLM testing [2], [3]. The key adaptation is that an LLM test is rarely a single exact-match assertion. It is often a dataset, rubric, grader, safety policy, tolerance threshold, and statistical decision rule that together define acceptable behavior.

Building on that foundation, this paper treats ATDLLMD as a systems engineering substrate for LLM application evaluation, not as a model benchmark alone. The proposed extension covers prompts, retrieval pipelines, tools, policies, model variants, logging, human review, monitoring, and governance documentation. It is business-centric because acceptance criteria are derived from stakeholder outcomes such as resolution rate, escalation burden, regulatory risk, response latency, support cost, and user trust. It is safety-centric because unacceptable harms are translated into durable tests and deployment gates. It is reliability-centric because repeated stochastic evaluation, regression baselines, and post-deployment drift monitoring are part of normal development rather than an afterthought.

The paper makes four contributions beyond the published ATDLLMD framing. First, it synthesizes a red-train-green lifecycle for acceptance-test-driven LLM applications. Second, it proposes a reference architecture that connects requirement contracts, evaluation harnesses, safety gates, data updates, and runtime monitoring. Third, it defines a metric stack spanning functional, safety, factuality, robustness, operational, and business outcomes. Fourth, it gives an empirical evaluation protocol that can be used to test whether ATDLLMD-style workflows improve real development outcomes compared with prompt-first or benchmark-after workflows.

# 2. Background and Related Work

## 2.1 From Software Tests to Behavioral Evaluation

Behavioral testing for NLP predates the current LLM wave. CheckList demonstrated that aggregate accuracy can miss important model failures and that structured behavioral tests help practitioners discover more defects [4]. HELM broadened language model evaluation by reporting multiple metrics across scenarios, including accuracy, calibration, robustness, fairness, bias, toxicity, and efficiency [5]. The Language Model Evaluation Harness made reusable few-shot evaluation infrastructure available to researchers and practitioners [6]. These efforts motivate ATDLLMD's view that evaluation must be engineered as a durable artifact, not improvised at the end of a project.

LLM-as-judge methods can scale semantic evaluation for open-ended outputs, but they introduce judge bias, positional effects, verbosity preference, and reasoning limits [7]. ATDLLMD therefore treats model-based grading as one evaluator among several. Deterministic validators, reference-based checks, human adjudication, adversarial probes, and statistical repeat runs are all needed when outputs affect high-stakes workflows.

## 2.2 Safety, Factuality, and Governance

SafetyBench and red-teaming research show that safety evaluation requires broad taxonomies and adversarial search rather than a few hand-written prompts [8], [9]. Factual reliability work similarly shows that simple accuracy is not enough when prompt and context variation can change whether a model gives a correct answer [10]. Citation-generation studies expose a concrete operational failure mode: plausible references and DOIs may be fabricated even when the surrounding prose appears scholarly [11]. These findings support the ATDLLMD rule that any escaped failure should become a permanent regression case.

The governance context is also maturing. The NIST AI Risk Management Framework frames trustworthy AI around validity, reliability, safety, resilience, accountability, transparency, privacy, and fairness [14], while the NIST generative AI profile adds risks specific to content generation and model interaction [15]. OWASP's LLM application taxonomy highlights prompt injection, sensitive information disclosure, supply-chain weakness, data and model poisoning, excessive agency, vector-store weakness, misinformation, and unbounded consumption [16]. ISO/IEC 42001 defines an organizational AI management system for responsible development and use [17]. ATDLLMD operationalizes these governance ideas by converting them into measurable acceptance gates.

## 2.3 Data and Metadata Discipline

LLM evaluation data is itself a product. Evaluation suites must preserve metadata about task type, source, risk level, language, user segment, retrieval corpus version, model version, prompt variant, and adjudication history. Work on efficient representation of high-cardinality categorical variables is relevant when eval systems need compact but expressive encodings for thousands of domains, intents, policies, and failure categories [22]. Metadata harmonization for language resources is similarly relevant because an acceptance-test registry becomes useful only when tests can be queried, compared, reused, and audited across projects [23].

Robustness analysis can also borrow from formal and statistical ideas outside LLM evaluation. Certified-radius estimation for randomized smoothing illustrates how uncertainty-aware robustness claims require explicit estimation assumptions, not just anecdotal perturbation tests [24]. Even mundane data-quality automation matters: date-format detection, for example, becomes important when operational dashboards, monitoring windows, audit timelines, and incident reports must be merged without hidden parsing errors [25]. These adjacent data-engineering concerns are part of a serious test-driven LLM lifecycle.

## 3. ATDLLMD Extension Framework

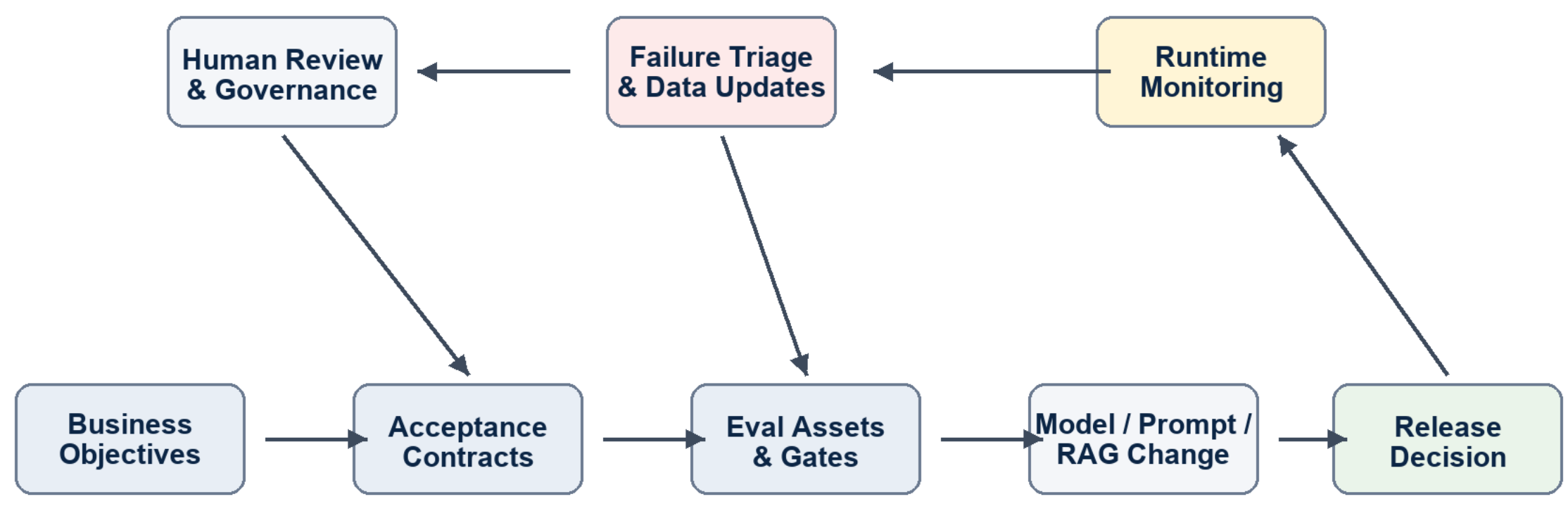

*Figure 1. ATDLLMD lifecycle. Acceptance contracts and evaluation assets precede model changes; runtime failures feed the next specification cycle.*

### 3.1 Core Principle

ATDLLMD begins with the claim that a deployable LLM feature should not start from a prompt. It should start from acceptance contracts: structured statements of what the system must do, what it must not do, how evidence will be measured, and who is accountable for residual risk. A contract may specify that a benefits assistant must answer policy questions only from the approved corpus, cite supporting sections, refuse requests for personal data, escalate ambiguous eligibility cases, complete responses within a latency budget, and maintain a target containment rate without increasing complaint risk.

The development loop is red-train-green. Red means the team first creates tests that the current system cannot pass or cannot pass reliably. Train is shorthand for all improvement mechanisms: prompt revision, retrieval tuning, tool restriction, policy update, fine-tuning, synthetic data generation, human labeling, or UI redesign. Green means the system satisfies its release gates with enough statistical confidence and traceability to justify deployment. The loop then continues in production as monitoring discovers new failures.

### 3.2 Lifecycle Phases

1. Business and risk framing. Identify stakeholder goals, unacceptable outcomes, regulatory constraints, operational budgets, and user groups.
2. Acceptance contract design. Translate goals into measurable functional, safety, factuality, robustness, cost, and governance criteria.
3. Evaluation asset construction. Build datasets, adversarial prompts, rubrics, deterministic validators, human review instructions, and traceability metadata.
4. System change. Modify prompts, retrieval, tools, policies, model choice, fine-tuning data, interfaces, or fallback workflows.

5. Automated and human validation. Execute the test suite repeatedly, compare with baselines, estimate confidence intervals, and adjudicate ambiguous cases.
6. Release and monitoring. Deploy only when gates pass; monitor drift, cost, user behavior, incident reports, and new prompt-injection or hallucination patterns.
7. Failure conversion. Convert production failures and review findings into durable tests, updated datasets, and revised contracts.

### 3.3 Reference Architecture

A practical ATDLLMD implementation separates the application runtime from the evaluation control plane. The runtime contains the user interface, orchestration layer, retrieval system, model endpoints, tool connectors, guardrails, logging, and escalation path. The control plane contains the requirement registry, evaluation datasets, model/prompt/version registry, scoring harness, statistical analyzer, release gate service, audit report generator, and monitoring feedback processor. This separation makes it possible to change an application while preserving historical comparability.

| Layer | Responsibility | Examples of Evidence |
|---|---|---|
| Requirement registry | Maps stakeholder goals to acceptance contracts and risk categories. | Requirement IDs, owner, risk severity, acceptance threshold, review date. |
| Eval asset store | Stores reusable datasets, adversarial suites, rubrics, and ground truth. | Prompt sets, reference answers, policy snippets, judge prompts, human labels. |
| Scoring harness | Runs candidate systems across deterministic, model-based, and human-in-loop evaluators. | Pass/fail results, confidence intervals, disagreement rates, seed/version logs. |
| Release gate | Blocks deployment when any critical contract fails or regresses. | Gate decision, waiver record, approver, residual-risk note. |
| Runtime monitor | Detects drift, incidents, unsafe outputs, cost spikes, and business-metric degradation. | Trace samples, escalation rate, complaint tags, latency, token spend, retrieval misses. |

*Table 1. ATDLLMD control-plane components and the evidence they preserve for development and audit.*

## 4. Metric Stack

ATDLLMD uses multidimensional gates because a single scalar score invites hidden trade-offs. A customer-support assistant might improve answer completeness while leaking sensitive data; a medical intake bot might reduce escalations while mishandling uncertainty; a coding assistant might pass functional tests while introducing insecure patterns. The framework therefore evaluates candidates along six dimensions.

| Dimension | Representative Measures | Typical Gate |
|---|---|---|
| Functional utility | Task success, exact constraints satisfied, response completeness, tool-call correctness. | No critical workflow below baseline; target task success exceeded. |
| Factual grounding | Supported-claim ratio, citation validity, retrieval attribution, contradiction rate. | All high-risk factual claims grounded in approved sources. |
| Safety and security | Unsafe compliance, false refusal, prompt-injection success, data leakage, excessive agency. | Zero critical safety failures in release suite; monitored residual risk. |
| Robustness | Paraphrase stability, multilingual consistency, noisy-context tolerance, regression from prior version. | No statistically significant regression in protected slices. |
| Operational quality | Latency, token cost, timeout rate, cache hit rate, evaluator cost. | Cost and P95 latency inside service budget. |
| Business value | Containment, escalation quality, user satisfaction proxy, complaint rate, conversion or resolution. | Business KPI improves without safety or quality regression. |

Table 2. ATDLLMD metric stack. Gates should be tailored to domain risk rather than copied across applications.

The metric stack should be versioned alongside prompts and code. Threshold changes require justification because changing a gate can be as consequential as changing a model. In high-risk settings, a release report should include dataset version, candidate system version, baseline comparison, statistical method, known blind spots, and waiver decisions.

# 5. Evaluation Protocol

## 5.1 Comparative Study Design

To test whether ATDLLMD provides value beyond good intentions, future work should compare it with two common baselines. The first is prompt-first iteration, in which developers manually adjust prompts after informal inspection. The second is benchmark-after development, in which a system is built first and then evaluated against a fixed benchmark. ATDLLMD differs by requiring acceptance contracts and failing tests before system changes.

A controlled study can use the same application, team size, time budget, model choices, and corpus across all workflows. Candidate domains include enterprise policy question answering, healthcare appointment triage, financial document summarization, legal clause extraction, and technical support. Each workflow should produce a release candidate after the same number of development cycles. A holdout evaluation suite, including production-like prompts not visible during development, can then compare task success, safety failures, hallucination rate, robustness, cost, and business proxies.

## 5.2 Statistical Treatment

Because LLM outputs vary across runs, evaluation should measure distributions rather than isolated examples. Repeated sampling is needed for generative tasks, and confidence intervals should be reported for key metrics. McNemar-style paired tests, bootstrap intervals, or Bayesian hierarchical models can compare candidate systems while accounting for prompt-level variance. Rare safety failures need separate treatment because a suite with no observed critical failures does not prove that the failure probability is zero.

Human adjudication should be calibrated. Annotators need written rubrics, examples, disagreement resolution, and periodic blind overlap. LLM-as-judge can reduce cost for low-risk semantic scoring, but its outputs should be audited against human labels and protected from prompt leakage or rubric manipulation [7].

# 6. Running Example: Regulated Support Assistant

Consider a regulated enterprise support assistant that answers employee policy questions using a retrieval-augmented generation architecture. The business wants to reduce help-desk load, but the assistant must avoid giving incorrect eligibility advice, exposing personal data, or fabricating policy citations. A prompt-first team might tune a system prompt until examples look good. Under ATDLLMD, the team first writes acceptance contracts.

- Functional contract: answer policy questions using retrieved policy sections and ask a clarifying question when required inputs are missing.
- Factual contract: every eligibility statement must be supported by an approved policy source and include a traceable citation.
- Safety contract: refuse requests to reveal another employee's protected information and ignore instructions embedded in retrieved documents.
- Business contract: reduce Tier-1 escalations while keeping complaint and re-open rates below the existing baseline.

- Operational contract: keep P95 response latency and per-resolution token cost within the service budget.

Development then begins with failing tests. The eval suite includes happy-path questions, ambiguous cases, adversarial prompt-injection documents, paraphrased questions, policy-version conflicts, and unsupported requests. If a candidate improves containment but increases unsupported citations, the release gate fails. If production monitoring later finds a new prompt-injection pattern, that trace becomes a permanent regression test. The system improves through repeated contract updates rather than through isolated prompt tinkering.

# 7. Discussion

## 7.1 Why ATDLLMD Is Not Just LLMOps

LLMOps often emphasizes deployment, monitoring, prompt management, model routing, and observability. ATDLLMD overlaps with those practices, but its distinguishing commitment is specification before implementation. Evals are not merely a dashboard; they are the mechanism by which business and safety requirements shape architecture. This distinction matters because many LLM failures are discovered only after users encounter them. ATDLLMD attempts to move those discoveries earlier and to convert unavoidable discoveries into permanent learning.

## 7.2 Limits and Risks

ATDLLMD does not eliminate uncertainty. Acceptance tests can encode incomplete requirements, flawed rubrics, biased data, unrealistic prompts, or the wrong business incentives. Overfitting to an evaluation suite is possible, especially when tests are static or too close to development examples. Safety gates can also create false confidence if they do not represent realistic attack paths. For these reasons, the framework requires rotating holdout sets, adversarial test generation, human review, production sampling, and explicit documentation of blind spots.

Another limitation is organizational cost. Stakeholders must participate in contract design, evaluators must be maintained, and data governance must support versioned traceability. However, these costs should be compared with the cost of post-deployment incidents, unrepeatable model-selection decisions, audit failure, or repeated manual review. In regulated and high-value settings, evaluation discipline is infrastructure.

# 8. Conclusion

This paper extends the ATDLLMD line of work by specifying evaluation protocols, release gates, and governance evidence for business-centric LLM systems. By adapting test-driven and acceptance-test-driven principles to probabilistic systems, the extension connects business value, safety, factual grounding, operational reliability, and governance into one lifecycle. Its practical promise is not that every LLM behavior becomes deterministic, but that development decisions become traceable, regressions become visible, and failures become reusable evidence. The next step is rigorous empirical comparison across real projects to quantify whether ATDLLMD-style workflows reduce escaped defects, improve stakeholder alignment, and lower the cost of safe iteration.